\documentclass{imsart}

\RequirePackage{amsthm,amsmath,amsfonts,amssymb}
\RequirePackage[authoryear]{natbib}
\RequirePackage[colorlinks,citecolor=blue,urlcolor=blue]{hyperref}
\RequirePackage{graphicx}
\usepackage{appendix}
\startlocaldefs
\theoremstyle{plain}

\newtheorem{theorem}{Theorem}[section]

\theoremstyle{remark}
\newtheorem{definition}[theorem]{Definition}
\newtheorem{property}[theorem]{Property}

\DeclareMathOperator*{\argmin}{\arg\!\min}

\endlocaldefs

\begin{document}

\begin{frontmatter}
\title{Tomographic reconstruction of a disease transmission landscape via GPS recorded random paths}
\runtitle{GPS-based tomography of disease transmission}

\begin{aug}
\author[A]{\fnms{Jairo}~\snm{Diaz-Rodriguez}\ead[label=e1]{jdiazrod@yorku.ca}},
\author[B]{\fnms{Juan Pablo}~\snm{Gomez}\ead[label=e2]{echeverrip@uninorte.edu.co}},
\author[C, D]{\fnms{Jeremy P.}~\snm{Orange}\ead[label=e3]{jporange2@ufl.edu}},
\author[E]{\fnms{Nathan D.}~\snm{Burkett-Cadena}\ead[label=e4]{nburkettcadena@ufl.edu}},
\author[F]{\fnms{Samantha M.}~\snm{Wisely}\ead[label=e5]{wisely@ufl.edu}},
\author[C, D]{\fnms{Jason K.}~\snm{Blackburn}\ead[label=e6]{jkblackburn@ufl.edu}}
\and
\author[G]{\fnms{Sylvain}~\snm{Sardy}\ead[label=e7]{sylvain.sardy@unige.ch}}

\address[A]{Department of Mathematics and Statistics,
York University, Canada\printead[presep={,\ }]{e1}}

\address[B]{Departamento de Qu\'imica y Biolog\'ia, Universidad del Norte, Colombia\printead[presep={,\ }]{e2}}

\address[C]{Spatial Epidemiology \& Ecology Research Laboratory, Department of Geography, University of Florida, USA\printead[presep={,\ }]{e3}}

\address[D]{Emerging Pathogens Institute, University of Florida, USA\printead[presep={,\ }]{e6}}

\address[E]{Florida Medical Entomology Laboratory, University of Florida, USA\printead[presep={,\ }]{e4}}

\address[F]{Department of Wildlife Ecology and Conservation, University of Florida, USA\printead[presep={,\ }]{e5}}

\address[G]{Mathematics Section, Universit\'e de Gen\`eve, Switzerland\printead[presep={,\ }]{e7}}
\end{aug}

\begin{abstract}
Identifying areas in a landscape where individuals have a higher likelihood of disease infection is key to managing diseases. Unlike conventional methods relying on ecological assumptions, we perform a novel epidemiological tomography for the estimation of landscape propensity to disease infection, using GPS animal tracks in a manner analogous to tomographic techniques in positron emission tomography (PET). Treating tracking data as random Radon transforms, we analyze Cervid movements in a game preserve, paired with antibody levels for epizootic hemorrhagic disease virus (EHDV)—a vector-borne disease transmitted by biting midges.
After discretizing the field and building the regression matrix of the time spent by each deer (row) at each point of the lattice (column), we model the binary response (infected or not) as a binomial linear inverse problem where spatial coherence is enforced with a total variation regularization. The smoothness of the reconstructed propensity map is selected by the quantile universal threshold. 
To address limitations of small sample sizes and evaluate significance of our estimates, we quantify uncertainty using a bootstrap-based data augmentation procedure. Our method outperforms alternative ones when using simulated and real data. This tomographic framework is novel, with no established statistical methods tailored for such data. 
\end{abstract}


\end{frontmatter}

\section{Introduction}


\subsection{Epidemiological background}
Predicting spatial disease risk is critical for managing infectious diseases that affect humans, livestock, and wildlife \citep{albery2022,miller2013diseases, Morris2016}. Although this area has historically been underexplored, recent advances in computational methods, animal tracking technologies, and modeling approaches have significantly improved our ability to infer the spatial dynamics of disease transmission \citep{dougherty2018going, white2018disease,manlove2022defining,dougherty2022framework}. Traditionally, research has focused on directly transmitted diseases, despite the fact that many major outbreaks and emerging zoonoses—such as anthrax, brucellosis, and vector-borne infections—are transmitted indirectly through environmental reservoirs or vectors \citep{altizer2011animal,rayl2021elk,dougherty2022framework}. Spatial predictions for such indirectly transmitted diseases often rely on aggregated case data or prior ecological knowledge. However, these approaches may be inadequate for emerging or rapidly evolving disease systems \citep{altizer2011animal, manlove2022defining}.

Mechanistic and spatially explicit compartmental models offer more formal approaches, but typically require system-specific information, such as transmission rates or assumptions about host-environment interactions, which may be unavailable for novel pathogens \citep{dougherty2018going,roosa2019assessing,glennon2021challenges,dankwa2022structural}. Consequently, understanding how wildlife moves across landscapes and interacts with environmental reservoirs or vectors has become increasingly important for identifying areas of elevated disease risk \citep{manlove2022defining}.

Modern species distribution modeling methods—such as Maxent, GARP, and boosted regression trees—have been used to associate disease risk with environmental or climatic covariates \citep{Elith2011,Ahmed2015,solano2019malaria,dougherty2022framework}. However, these presence-only models rarely account for host movement dynamics and behavioral heterogeneity \citep{dougherty2018going,dougherty2022framework}. Incorporating home range analyses and GPS-based movement data has proven effective for mapping zoonotic disease hotspots and clarifying transmission routes \citep{rayl2021elk}. Linking infection status with high-resolution movement data—such as that derived from telemetry—enables researchers to identify high-risk spatial zones and critical time windows for pathogen transmission \citep{dougherty2018going,wilber2022model,malmberg2025cross}. Nevertheless, this approach remains underutilized in disease niche modeling.

Recent studies have increasingly integrated animal movement data to identify areas of elevated infection risk and support targeted surveillance and intervention strategies \citep{dekelaita2023animal,ciss2023description,mcduie2024mitigating}. While spatially explicit compartmental models have improved risk prediction \citep{white2018disease}, these approaches typically require a priori knowledge of disease ecology or model-specific parameters \citep{manlove2022defining,gao2023model,akter2025conditional} . Other models rely on detailed information about interactions between infected and susceptible individuals \citep{wilber2022model}. While these frameworks represent major advances, they still depend on assumptions about disease transmission pathways and host-environment relationships.

In this study, we present and validate a novel method for mapping infectious disease risk that combines individual-level movement trajectories and health status data without requiring prior knowledge of the transmission process or environmental context. This method can serve as a powerful stand-alone tool or be integrated with environmental datasets to link spatial risk predictions with ecological mechanisms of disease transmission \citep{manlove2022defining}. We demonstrate the utility of this approach using both simulated data and a case study of native and exotic cervids infected with epizootic hemorrhagic disease virus (EHDV) on a Florida hunting ranch.

\subsection{Cervid movement and EHDV prevalence data collection} \label{subsct:data}

In a 172-ha high-fenced ranch in Florida, USA dedicated to big-game hunting, move freely  white-tailed (WTD), Pere David (ED), Fallow (DD) and Elk deer (CC). A high prevalence of several serotypes of  EHDV  causes severe clinical signs such as hemorrhaging, edema, hoof-sloughing, oral lesions and death, principally in WTD, but it affects other cervid species as well. 
EHDV is transmitted by biting midges from the genus \textit{Culicoides} and, in southeastern United States, \textit{Culicoides stellifer} and \textit{Culicoides venustus} have been identified as the competent EHDV vectors. There has been substantial research on this high-fenced ranch regarding host behavior, spatial movement and vector behavior and ecology  \citep{mcgregor2019host, dinh2021midge,mcgregor2019field,orange2021homerange,dinh2020TLoCoH, dinh2021resource}. The prevalence of EHDV in WTD and other exotic species has also been described \citep{cauvin2020, orange2021exotic}, but to date, disease risk on the ranch has been estimated by inferring resource selection in deer and overlap with areas of high biting midge abundance. 

To investigate disease transmission, we used data from 26 GPS collared WTD (8 females, 18 males) tracked for an average of 149 days during 2015 and 2016. Individuals were captured and GPS collared in the spring, ahead of the EHDV transmission season \citep{dinh2020TLoCoH} and recaptured in fall to remove the collars.  GPS collars were programmed to collect a GPS location every 15 or 120 minutes \citep{cauvin2020,dinh2020TLoCoH}. Animals were immobilized with cartridge or air powered darts, monitored, collared, bled, and released (see \citealp{cauvin2020,dinh2020TLoCoH} for details about capture procedures). A blood sample was obtained from individual at the time of initial capture and again when GPS collars were recovered. Each blood sample was tested for the presence of EHDV-1, EHDV-2 and EHDV-6 antibody titers using a virus neutralization test performed at Texas A\&M University (\citealp{stallknecht1996}; see \citealp{cauvin2020} for details). We classified animals as positive or negative for each virus type using previous cutoffs (\citealp{cauvin2020}). Four individuals were tracked during both years allowing us to use the tracks as independent replicates to increase our sample size to 30 animals (See Supplementary materials). We pre-processed the raw GPS tracks and excluded those of seropositive individuals at both collection times. All tracks were interpolated to 15 minutes. At the end the data consisted of 27 tracks for EHDV-1 infected animals, 22 for EHDV-2, and 30 for EHDV-6. The deer capture and handling protocols were developed by JKB and the ranch wildlife manager and approved by the Institutional Animal Care and Use Committee at the University of Florida (UF IACUC Protocols \#201508838 and \#201609412).

\subsection{Our contributions}
We propose a novel framework to estimate the landscape propensity of EHDV infection using only deer movement data, without relying on external environmental or ecological covariates. Individual deer wandering across the landscape are conceptualized as tomographic rays traversing a medium, while the corresponding binary infection outcomes serve as sensor readings (Figure \ref{fig:schema}). We regularize the high dimensional inverse problem with a total-variation approach and derive the mathematical and practical guidelines for choosing the penalty parameter, yielding models that remain readily interpretable for practitioners.

In Section \ref{sct:datat} we describe a data-transformation procedure, inspired by the Radon transform in positron-emission tomography (PET), that recasts the task as an epidemiological tomographic inverse problem. Because the number of GPS-collared deer is typically small, this transformation produces a high-dimensional inference problem with limited observations. In Section~\ref{sct:TV}, we model the  data with a generalized linear model regularized by a total variation penalty to enforce spatial coherence in the propensity map. In this front, we derive the necessary mathematical result for an efficient selection of the regularization parameter with the quantile universal threshold,  that does not require cross-validation. We propose an algorithm to solve the convex optimization problem, we derive a statistical test for a constant propensity map, and we show how to perform uncertainty quantification.
In Section~\ref{sct:alternative}, we explore alternative tomographic approaches, and we also introduce a potential subsequent data aggregation designed to accommodate direct image smoothing methods.
In Section~\ref{sct:MC}, we perform a Monte Carlo simulation to empirically compare all methods with three different asymptotics: increasing sample size, increasing spatial resolution, and increasing frequency recording.
In Section~\ref{sct:appli}, we analyze the deer data of Section~\ref{subsct:data}.
In Section~\ref{sct:conclu}, we summarize our contributions and point to generalizations of our approach.
The research is reproducible (see Supplementary Materials).



\section{Tomographic data transformation}
\label{sct:datat}

\begin{figure}[!t]
\centering
\includegraphics[width=5.3in]{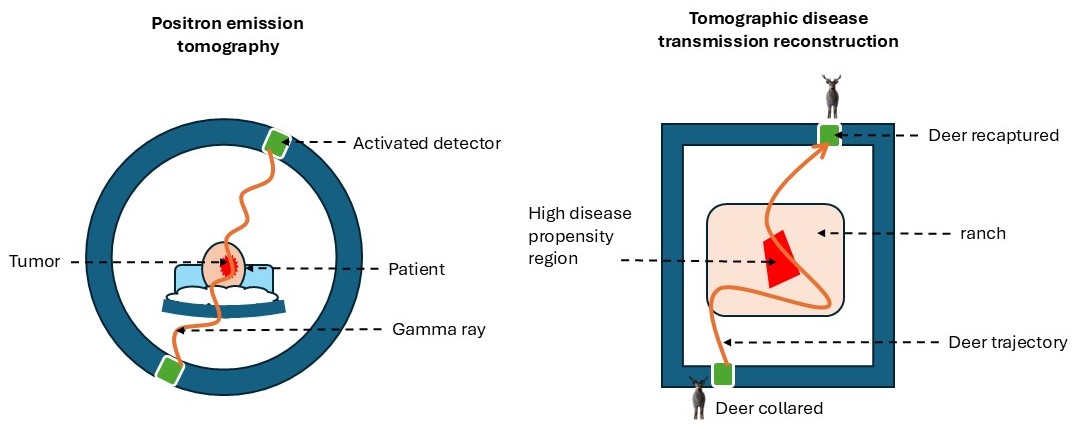}
\caption{An analogy between Positron Emission Tomography (left) and the tomographic reconstruction of a disease propensity map (right): Deer moving in a ranch represent gamma rays traveling through a patient. Whether or not a deer gets infected is like a positron is emitted in PET. The goal is to reconstruct the high-propensity area, similar to identifying a tumor.}
\label{fig:schema}
\end{figure}

If, for a second, we imagine that the deer described in Section~\ref{subsct:data} could be trained to enter the ranch at designed locations, walk along lines $L_i$ and exits the ranch at the end of their straight walks, then the information whether the deer contracted the disease along these lines could help map the disease propensity on the ranch and each deer walking along a line would essentially have calculated the Radon transform $R_\mu(L_i)=\int_{L_i} \mu(u,v) \sqrt{du^2+dv^2}$ of the propensity map $\mu(u,v)$ as a function of locations $(u,v)$, along the $L_i$, for deer and line $i=1,\ldots,n$. This is what positron emission tomography (PET) is doing with rays instead of deer, for non-invasive brain imaging for instance (Figure~\ref{fig:schema}).

This is not feasible with deer because they do not move along straight lines but, while moving freely along random paths $P_i$, deer gather information about $\mu(u,v)$ with the Radon transform $R_\mu(P_i)=\int_{P_i} \mu(u,v) \sqrt{du^2+dv^2}$, for deer and path $i=1,\ldots,n$. If one can extract information from GPS locations of individuals moving freely in a medium along random paths and contracting a disease or not, then one performs some imaging of the medium, here the disease propensity map.

Going back to the deer movement data, each deer has a blood test at time $0$ leading to concentrations of four different antibodies $c_{i,m}(0)$, and at time $t_i$ leading to concentrations $c_{i,m}(t_i)$, $m=1,\ldots,3$ (EHDV-1, EHDV-2 and EHDV-6 antibodies). From these concentrations, 
the binary responses $y_{i,m}=1_{\{c_{i,m}(t_i)>c_{i,m}(0)\}}$
declares whether or not the $i$th individual got infected by virus $m$ during the time period,  for $i=1,\ldots,n$ and $m=1,\ldots,3$.
Individuals move differently in the fenced area~$\Omega$, as observed by the GPS tracking system.
To map the spatial variability in infection propensity $\mu$, the fenced area~$\Omega$ is spatially descretized and partitioned over equal small units $\omega^{j,k}$ indexed by latitude $j$ and longitude $k$, for $(j,k)\in {\cal F}$. The level of discretization $p=|{\cal F}|$ is chosen by the practitioner: the larger $p$, the finer the spatial discretization, but the more parameters to estimate.
Correspondingly, each area $\omega^{j,k}$  has the propensity $\mu(\omega^{j,k})$ of contaminating a species with a disease during a unit time period, where $\mu$ is a function of localization $\omega^{j,k}$ in $\Omega$. Using the GPS recordings, one can, for each individual $i$, aggregate the total time $x^{j,k}_{i}$ spent in each small area $\omega^{j,k}$ between its starting time until time $t_i$; so let ${\bf x}_{i}=\{x_i^{j,k}\}_{(j,k)\in {\cal F}}$ be the vector of total time spent in location $\omega^{j,k}$ for individual $i$, for $i=1,\ldots,n$.
The goal is to estimate the propensity maps ${\boldsymbol \mu}_m=\{\mu_m(\omega^{j,k})\}_{(j,k)\in {\cal F}}$ ($m=1,\ldots, 3$, one map per EHDV virus) based on the information contained in the data $(y_i, { \bf x}_i)_{i=1,\ldots,n}$. We performed this data transformation process (we provide the resultant $X$ and $y$ for each dataset in the Supplementary materials) to the EHDV-1, EHDV-2 and EHDV-6 data.
Individuals randomly moving in the field resemble tomographic rays passing through a medium, with their binary infection outcomes acting like sensor measurements (Figure~\ref{fig:schema}).
So we aim at retrieving the disease propensity map, in the spirit of inverse problems such as positron emission tomography (PET). 
To the best of our knowledge, our approach is novel to disease mapping and indirect risk estimation for the data we described.
\section{Regularization of GLM with total variation and the quantile universal threshold} \label{sct:TV}

\subsection{Model}

Since the three EHDV serotype antibodies considered have their own specificities, three separate models are estimated, so we call the disease marker $y_i\in\{0,1\}$ for $i=1, \ldots, n$ (instead of $y_{i,1}$ for EHDV-1, $y_{i,2}$ for EHDV-2 and $y_{i,3}$ for EHDV-6).  
The classical approach for binary outcomes sees $y_{i}$ as a realization of the random variable $Y_{i}\sim \operatorname{Bernoulli}(\epsilon_{i})$, where the probability $\epsilon_i$ of being infected depends on where the $i$th individual spent its time in the fenced area. In particular $\epsilon_i$ should be high if the $i$th individual spent long periods of time in areas of high infection propensity; in other words, $\epsilon_i$ should be high when high total times~$x_i^{j,k}$ are observed in regions $\omega^{j,k}$ where the propensity $\mu(\omega^{j,k})$ is high.
A simple yet realistic model   is the generalized linear model \citep{NW72} which assumes  
$g(\epsilon_i)={\bf x}_i^{\rm T} {\boldsymbol \mu}$, where  the \emph{logit} link  $g(\epsilon)=\log(\epsilon/(1-\epsilon))$ maps $(0,1)$ into ${\mathbb R}$. Letting ${\bf Y}=(Y_1, \ldots, Y_n)$ and $X$ be the $n\times p$ matrix which $i$th row is ${\bf x}_i^{\rm T}$, the model is
\begin{equation}\label{eq:model}
{\bf Y} \sim \operatorname{Bernoulli}(g^{-1}(X{\boldsymbol \mu})),
\end{equation}
where ${\boldsymbol \mu}$ is the unknown propensity vector.
So the model can be seen as a tomographic linear inverse problem, for which the individuals are probing the space with their distinct movements reflected in the data matrix $X$ of total times visting the lattice system in $\Omega$. Instead of directly measuring the sources of where the virus is spreading, the information is indirectly measured on individuals living in potentially infecting areas.

%

\subsection{Estimation}

The vector of infection propensity  ${\boldsymbol \mu}$ has length $p=|{\cal F}|$, the cardinality of the lattice system that segments the fenced area $\Omega$. In our application the number of individuals is small, so we choose a fairly coarse segmentation  into $p=200$ cells. Regularization is needed, and owing to the spatial structure of the problem, smoothness can be imposed on ${\boldsymbol \mu}$ in many ways (see Section~\ref{sct:alternative}).
We choose total variation (TV) smoothing \citep{ROF92} for several reasons: first the domain is not rectangular, second epidemiologists prefer to identify piecewise-constant propensity regions 
and third the propensity map may have some peaks (e.g., feeders) and discontinuities (e.g., along a river or a lake). TV provides a solution to these challenges as it easily adapts to a non-regular lattice, it performs segmentation (as it fits a piecewise constant function) and it allows sudden peaks and jumps. We also propose a novel simple rule to select the smoothing parameter $\lambda$.

TV requires a system of neighborhood in the propensity map ${\boldsymbol \mu}$.
An entry $\mu_l$ of the vector of infection propensity ${\boldsymbol \mu}$ corresponds to a small region $\omega^{j,k}$ in~$\Omega$. Each region $(j,k)$ is associated to an entry of ${\boldsymbol \mu}$, call it $l$. Each region $(j,k)$ (associated to $l$) has neighbors: call $\partial l$ the set of entries of ${\boldsymbol \mu}$  corresponding to the neighbors of $(j,k)$. Because of the spatial structure of ${\boldsymbol \mu}$ which maps the propensity of catching a virus somewhere in the ranch, one believes that the value of $\mu_l$ is close to $\mu_{l'}$ for $l'\in \partial l$; here we consider the north-south-east-west neighborhood of a cell.
TV imposes  smoothness  by solving
\begin{equation}\label{eq:tv}
\min_{{\boldsymbol \mu}\in{\mathbb R}^p} -\log {\rm L} (g^{-1}(X{\boldsymbol \mu});{\bf y})+\lambda \sum_{l=1}^p \sum_{l' \in \partial l} |\mu_{l}-\mu_{l'}|,
\end{equation}
where $L$ is the likelihood associated to~\eqref{eq:model}, ${\bf y}$ is the vector of infection indicator per individual,  $X$ is total time matrix per individual (lines) spent in area $\omega^{j,k}$ (column), and ${\boldsymbol \mu}$ is the two-dimensional surface to reconstruct.
The regularization parameter $\lambda>0$ controls the smoothness of the estimate $\hat {\boldsymbol \mu}_\lambda$ solution to~\eqref{eq:tv}. One standard selection of $\lambda$ is cross-validation, which requires $n$ large for a stable estimation. In our situation, the number $n$ of individuals is rather small, so we employ the quantile universal threshold \citep{Giacoetal17} that is geared towards estimating ${\boldsymbol \mu}$ rather than good prediction of $X{\boldsymbol \mu}$. The quantile universal threshold is based on calibrating a choice of $\lambda$ under the null hypothesis $H_0$ that the propensity map is constant (that is, no regions are more infectious than others): its goal is to retrieve a constant map with high probability under $H_0$. This choice of $\lambda$ remains performant under alternative hypotheses as supported by the LASSO theory \citep{Tibs:regr:1996,BuhlGeer11} and total variation \citep{ROF92,sardy2019}, and demonstrates strong empirical performance in practical applications \citep{diaz2021nonparametric}.
Deriving the quantile universal threshold for our tomographic TV-estimator~\eqref{eq:tv} requires the following property.

\begin{property} \label{prop:lambda0}
Assuming the entries of the responses ${\bf y}$ are not all identical (all ones or all zeros),
there exists a finite $\lambda>0$ for which the solution $\hat {\boldsymbol \mu}_\lambda$ to~\eqref{eq:tv} is a constant propensity map across $\Omega$. 
Moreover the smallest such $\lambda$ is given by the zero-thresholding function
\begin{equation} \label{eq:ztf}
 \lambda_0({\bf y}, X)=\min_{{\boldsymbol \omega}\in {\mathbb R}^q} \| {\boldsymbol \omega}\|_\infty 
 \end{equation}
 $${\rm s.t.} \quad  X^{\rm T}({\boldsymbol \epsilon}-{\bf y})+D^{\rm T}{\boldsymbol \omega} ={\bf 0},
\ {\boldsymbol \epsilon}=g^{-1}(\hat \beta_0 X{\bf 1}),
\ \hat \beta_0=\argmin_{\beta_0\in {\mathbb R}} h(\beta_0 {\bf 1}; {\bf y}),$$
where $h({\boldsymbol \mu} ;{\bf y})=-\log {\rm L} (g^{-1}(X{\boldsymbol \mu});{\bf y})$ is the negative log-likelihood associated to~\eqref{eq:model},
and $D$ is the $q\times p$ matrix of the $q$ finite differences such that
$\| D{\boldsymbol \mu}\|_1=\sum_{l=1}^p \sum_{l' \in \partial l} |\mu_{l}-\mu_{l'}|$. 
\end{property}

The quantile universal threshold for the estimator~\eqref{eq:tv} can now be defined.

\begin{definition}[Quantile universal threshold] \label{def:qut}
Given a matrix $X$, consider the random variable
 $\Lambda=\lambda_0({\bf Y}, X)$,
 where ${\bf Y}$ is given by model~\eqref{eq:model} under the assumption that ${\boldsymbol \mu}$ is constant.
 For some small $\alpha\in(0,1)$, the quantile universal threshold $\lambda^{\rm QUT}_\alpha$ is the upper $\alpha$-quantile of $\Lambda$.
\end{definition}

Since $\alpha$ corresponds to the false discovery rate (of detecting a non-constant map) under $H_0$, one chooses $\alpha$ small, say $\alpha=0.05$. 
The distribution of $\Lambda$ is unknown however, so $\lambda^{\rm QUT}_\alpha=F^{-1}_\Lambda(1-\alpha)$ has no closed form expression. 
Instead $\lambda^{\rm QUT}_\alpha$ can be estimated by Monte Carlo, simulating $m$ realizations ${\bf y}^{(1)}, \ldots, {\bf y}^{(m)}$ of ${\bf Y}\sim \operatorname{Bernoulli}(g^{-1}(\beta_0 X{\bf 1}))$ for some $\beta_0$, for instance $\hat \beta_0=\argmin_{\beta_0\in {\mathbb R}} h(\beta_0 {\bf 1}; {\bf y})$ for the data ${\bf y}$ at hand. Then by calculating the corresponding $\lambda_0({\bf y}^{(1)}, X), \ldots, \lambda_0({\bf y}^{(m)}, X)$ and by taking their empirical upper $\alpha$-quantile, one estimates $\lambda^{\rm QUT}_\alpha$. The zero-thresholding function implicitly defined in Property~\ref{prop:lambda0} requires solving the non-trivial optimization problem in~\eqref{eq:ztf}; fortunately,  \eqref{eq:ztf} can be rewritten as an easily solvable linear program, namely
$$
\min_{{\boldsymbol \omega}, \lambda} ({\bf 0}^{\rm T}, 1)\left (
\begin{array}{c}
 {\boldsymbol \omega}\\
\lambda
\end{array}
\right )
 \quad {\rm s.t.} \quad
 \left \{
\begin{array}{c}
 {\bf u} \leq D^{\rm T} {\boldsymbol \omega} \leq {\bf u} \\
 {\boldsymbol \omega}-\lambda {\bf 1} \leq {\bf 0} \\
{\boldsymbol \omega}+\lambda {\bf 1} \geq {\bf 0} \\
\lambda \geq 0
\end{array}
\right . ,
$$
with ${\bf u}=X^{\rm T}({\bf y}-{\boldsymbol \epsilon})$, ${\boldsymbol \epsilon}=g^{-1}(\hat \beta_0 X{\bf 1})$ and
$\hat \beta_0=\argmin_{\beta_0\in {\mathbb R}} h(\beta_0 {\bf 1}; {\bf y})$.

\subsection{Testing} \label{subsct:TVtest}

Testing is part of statistical inference, and the TV regularization of the GLM~\eqref{eq:tv}  allows to do so.
A null hypothesis of interest is that there is no particular infectious region, in other words $H_0: \ {\boldsymbol \mu}=\mu_0 {\bf 1}$ is constant. When $n\geq p$, the likelihood ratio test (either based on the asymptotic $\chi^2$ distribution or on the exact distribution) comes to mind. 

When the number of recorded deer is small and the field's mapping is finely discretized, we are in the $p>n$ situation, however. To allow testing whether $n>p$ or not, we propose the TV-test~\citep{threshtest2022} and based on total variation smoothing~\citep{sardy2019}. The TV penalty term in~\eqref{eq:tv}  is  zero for the constant infectious propensity. Hence, given $\alpha\in(0,1)$, the test function
\begin{equation} \label{eq:TVtest}
\phi({\bf y}, X)= \left \{
\begin{array}{rl}
0 & {\rm if}\ \lambda_0({\bf y}, X)< \lambda^{\rm QUT}_\alpha \\
1 & {\rm otherwise.}
\end{array}
\right .
\end{equation}
 tests $H_0$ at level $\alpha$. One expects the TV-test to be more powerful than the likelihood ratio test for smooth (spatially homogeneous) alternative hypotheses~\citep{threshtest2022}. A Monte Carlo simulation confirms (See Supplementary materials). 
As a side product, when the test is not rejected for some level~$\alpha$ means that the solution to~\eqref{eq:tv} for $\lambda=\lambda^{\rm QUT}_\alpha$ is a constant propensity map without having to solve~\eqref{eq:tv}.

\subsection{Data augmentation with uncertainty quantification}
\label{subsct:uncertainty}

Estimation of the propensity  alone provides limited practical value for ecologists and practitioners.
Moreover GPS sampling constraints often result in small sample sizes $n$. 
To account for the uncertainty arising from the randomness of the observed independent individuals and their limited sample size, we employ the bootstrap that augments the sample size to $n_{\text{boot}}>n$ individuals.
To generate spatially correlated tracks so as to reflect the gregarious behavior typical of herds, we  sample the random moves for each resampled individual.
This bootstrapped data augmentation procedure is repeated multiple times, and the pointwise $(1-\alpha)$ quantiles of the resulting bootstrapped propensity estimates provide a measure of uncertainty. 

Also, due to the fact that the universal threshold $\lambda^{\rm QUT}_\alpha$ tends to be conservative, it shrinks and biases the total variation differences of the propensity estimates toward zero. To improve both the estimation accuracy and the coverage properties of the uncertainty intervals, we apply a pointwise bias correction $\hat {\boldsymbol \mu}^{\text{bc}} = 2\hat {\boldsymbol \mu} - \overline{\boldsymbol \mu}^{\text{boot}}$, where $\overline{\boldsymbol \mu}^{\text{boot}}$ represents the averaged bootstrapped propensity estimates \citep{CIS-2130}. We perform uncertainty quantification on the Cervid dataset in Section~\ref{sct:appli}.


\section{Alternative methods}\label{sct:alternative}
\subsection{Direct preprocessing}\label{sct:direct}

An alternative data aggregation strategy 
computes a naive empirical estimator of the infection propensity at each spatial location:
\begin{equation}\label{eq:empirical}
\hat \mu^{\rm emp}_m(\omega^{j,k}) = \frac{\sum_{\{i|y_i=1\}}
x_i^{j,k}}{\sum_{i=1}^n x_i^{j,k}}.
\end{equation}
This estimator can be interpreted as the empirical infection rate at each location, conditional on exposure. One may then smooth the log-odds transformation of this empirical rate:\begin{equation}
\log\left(\dfrac{\hat \mu^{\rm emp}_m(\omega^{j,k})}{1-\hat \mu^{\rm emp}_m(\omega^{j,k})}\right).
\end{equation}
Thanks to the structure induced by the aggregation, this approach enables the direct application of standard spatial smoothing methods.
\subsection{Smoothing methods}

Our data transformation casts the estimation problem as a tomographic linear inverse problem, that we regularized with total variation \citep{ROF92}. Whether we use the tomographic preprocessing data from Section~\ref{sct:datat} or the direct preprocessing data from Section~\ref{sct:direct}, 
a range of other regularization strategies can be employed, depending on prior assumptions about spatial smoothness or structure. For example, splines- or wavelets-based smoothing \citep{Wahb:spli:1990,SAT01} and Laplacian (Tikhonov) regularization \citep{zhou2003learning, tikhonov1963solution} have a smoothness penalty.
Spatially varying coefficients models \citep{gelfand2003spatial} allow the effect of covariates or exposure to vary continuously across space, capturing local heterogeneity more flexibly. Other options include alternative penalties to accommodate transitions \citep{tibshirani2005sparsity}, Gaussian Markov random fields \citep{rue2005gaussian} for probabilistic spatial smoothing, or low-rank basis expansions for dimensionality reduction \citep{cressie2008fixed}, and Gaussian Process Regression (GPR) or kriging \citep{matheron1963principles}.


\section{Monte Carlo simulation} \label{sct:MC}

\subsection{Estimation of disease propensity maps}\label{sec:simframework}

\begin{figure}[!t]
\centering
\includegraphics[width=4.0in, height = 1.2in]{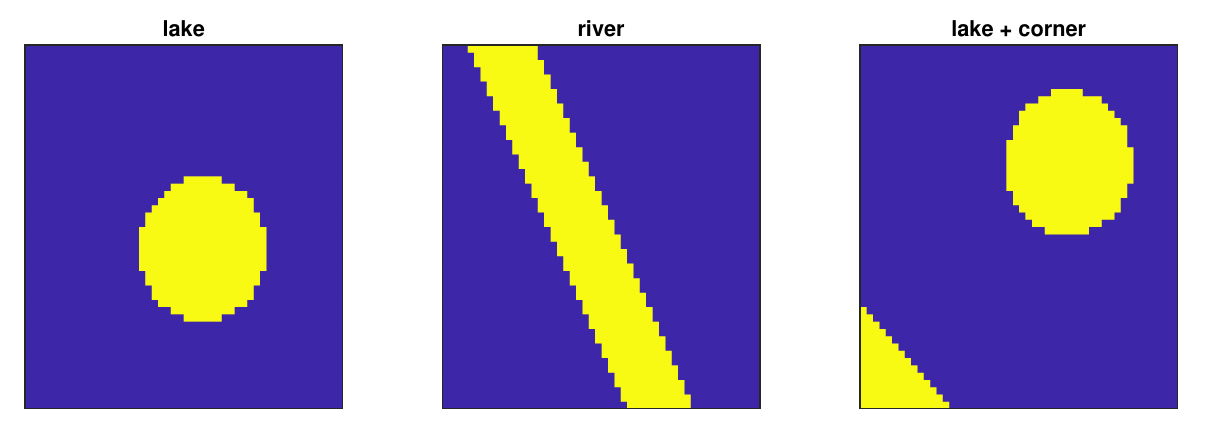}
\caption{Test functions for propensity map ${\boldsymbol \mu}_0$}
\label{fig:profiles}
\end{figure}

We generate an entire population of $n_0=5000$ individuals moving over a square fenced area $\Omega_0$ made of $N_0\times N_0$ lattice with known propensity 
${\boldsymbol \mu}_0$ given by one of the three binary profiles in Figure~\ref{fig:profiles} which simulates (from left to right) a lake, a river or a lake plus a corner.
We choose a full discretization with $N_0=50$ to provide a fine mapping of the area.
For individual $i\in\{1,\ldots, n_0 \}$, we simulate 
$T=2880$ random moves $l^{(i)}_1, l^{(i)}_2,\cdots,l^{(i)}_T$ on the lattice $\Omega$. 
The value of $T=2880$ mimics GPS samplings every $t=15$ minutes for 30 days.
At each time, the proposed moves are either to stay at the current location or to move to one of its four neighbors.
We simulate two different herd behaviors: one half of the individuals moves randomly with equal probability $1/5$ to one of the five proposed moves, while the other half moves with a probability that depends on ${\boldsymbol \mu}_0$ in a way that doubles the probability of moving to a region of higher virus infection propensity.
We call $L_0$ the $n_0\times T$ matrix which entry $(i,j)$ corresponds to the lattice locations of the $i$-th individual at time $t$. 
Based on $L_0$, we build the  $n_0\times p_0$ matrix~$X_0$ which counts the time spent in each of the $p_0=N_0^2$ lattice locations of $\Omega$ for each of the $n_0$ individuals. 
Then we generate  the binary infection vector ${\bf y}_0$, whether or not the animals were infected during the time period, according to model~\eqref{eq:model}. We expect more ones in the entries of ${\bf y}_0$ corresponding to the second herd which stays longer in regions of higher virus infection propensity.

To investigate the impact of the spatial discretization of $\Omega_0$, we  disctretize it into $(N\times N)$ lattices with  $N \in \{30, 50\}$, the largest $N$ corresponding to the full $\Omega_0$ plotted in  Figure~\ref{fig:profiles}. To investigate the impact of different frequencies of the tracers placed on the individuals, we sample the $L_0$ matrix every $t\in\{1,96\}$ units, a unit being 15 minutes, which means that each tracer sends its current location either every 15 minutes or  24 hours. The corresponding matrix $L_t$ of GPS locations built from  taking every $t$ columns of $L_0$ has $2880$, or $720$ columns, leading to a regression matrix $X$ of size $n_0 \times p$, where $p=N^2$.   Finally the number of tracers placed on the full population, that is, the number of observations, is taken as $n\in\{500, 5000\}$, the largest $n$ corresponding to the full population $n_0=5000$.


For increasing sample sizes  $n\in\{500, 5000\}$, 
increasing spatial discretization   $N \in \{30, 50\}$,
increasing temporal frequencies $1/t$ with $t\in\{96, 1\}$,
we perform Monte Carlo simulations for each combination of the three propensity maps ${\boldsymbol \mu}_0$ plotted in Figure~\ref{fig:profiles}. Each scenario is simulated 1000 times, leading to 1000  full data information triplets $({\bf y}_0, L_0, X_0)$, from which a response vector ${\bf y}$ is extracted and a regression matrix $X$ is built, depending on the value of $(n, N, t)$. For each of the 1000 data sets generated,  we estimate the propensity map with our tomographic TV solution to~\eqref{eq:tv} with the quantile universal threshold of Definition~\ref{def:qut} for the value of $\lambda$. For comparison we also estimate the propensities using four smoothing approaches applied to our tomographic preprocessing data from Section~\ref{sct:datat} and four smoothing approaches applied to the alternative direct preprocessing data from Section~\ref{sct:direct}. For the tomographic data we use Laplacian regularization with cross-validation, and splines and spatially variable coefficients (SVC) using \texttt{mgcv} R package \citep{wood2001mgcv}. For the direct data we compare with the naive empirical estimate, and its smoothed versions using GPR with cross-validation, and total variation with \texttt{tvR} R package \citep{tvR}) and Laplacian smoothing with regularization parameter  $\lambda = \hat{\sigma}\log{p}$. Spatially varying coefficients and splines can be seen as equivalent to GPR for the direct preprocessing data.

Owing to the different values of $(n, N, t)$, we scale all estimates on the interval $[0,1]$ to compare them in terms of mean squared error calculated as ${\rm MSE}=\sum_{l=1}^p (\hat \mu_l - \tilde \mu_{0,l} )^2/p$, where $\tilde {\boldsymbol \mu}_{0} $ is the nearest-neighbor interpolation of ${\boldsymbol \mu}_{0}$ on a coarser scale and $p=N^2$.

\begin{figure}[!t]
\centering
\includegraphics[width=5in]{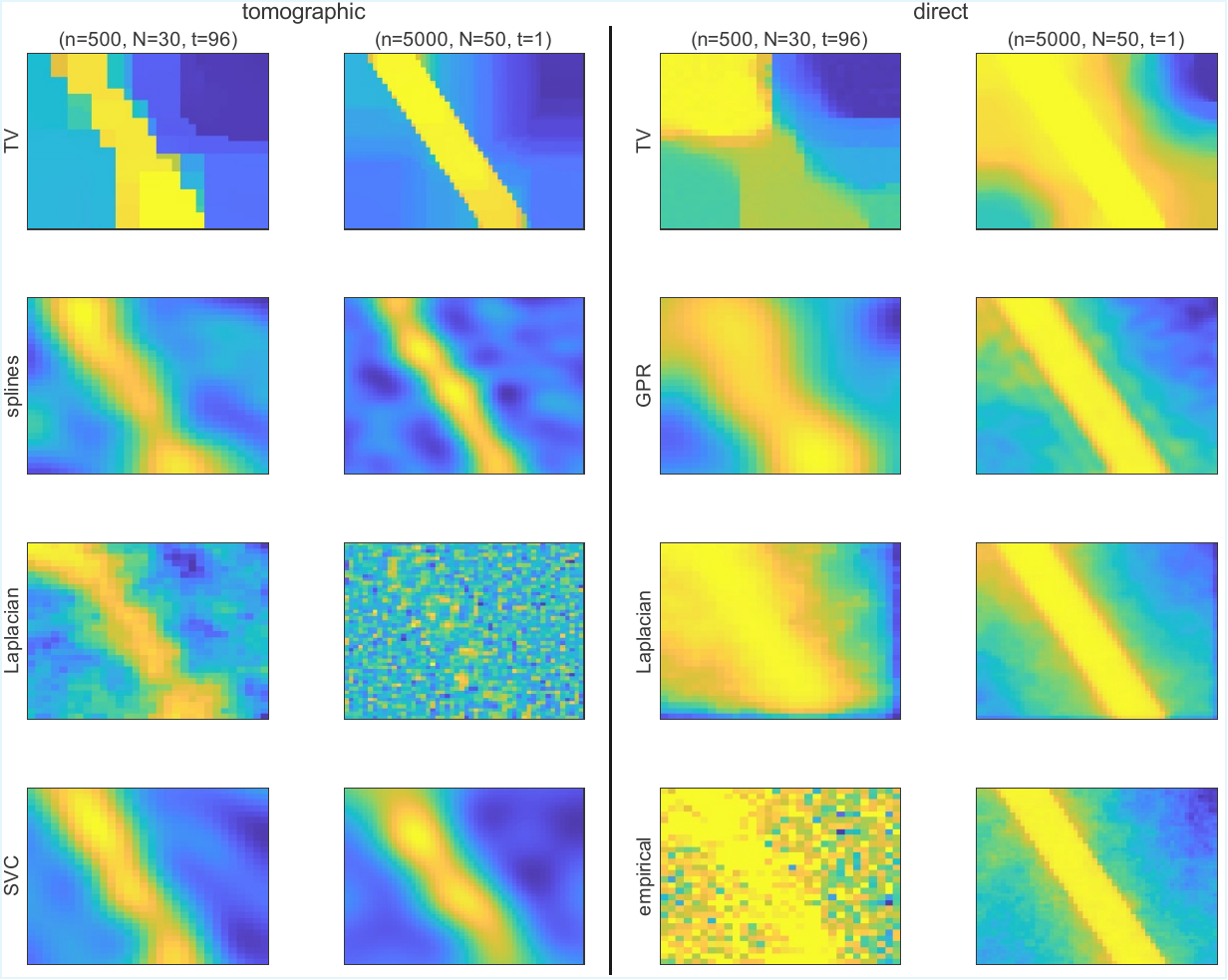}
\caption{Performance based on increasing $(n, N, t)$ from $(500, 30, 96)$ to $(5000, 50, 1)$ for all tomographic smoothing estimators (left) and direct smoothing estimators (right), for  \texttt{river} propensity map.}
\label{fig:asymptotics}
\end{figure}

Table~\ref{meanMSE} reports the  MSE between $\hat {\boldsymbol \mu}$ and $\tilde {\boldsymbol \mu}_0$ averaged over the 1000 Monte Carlo runs.
Figure~\ref{fig:asymptotics} shows typical estimations for the \texttt{river} propensity map. 
Our tomographic TV method outperforms in nearly all scenarios.
Among the tomographic estimators, spline‑based and SVC reconstructions consistently underestimate the magnitude of high‑propensity regions, whereas the Laplacian method fails to detect these areas altogether. Although the direct smoothing estimators appear slightly crisper with larger sample sizes, they still yield the highest mean‑squared error and produce increasingly noisy propensities.
As expected, MSE decreases for the asymptotic we considered, namely
when the sample size $n$ of individuals increases, the spatial discretization $N$ increases, and the GPS frequency $1/t$ increases.



\begin{table}[!htbp]
  \caption{Results of Monte Carlo simulation reporting the averaged mean squared errors between $\hat {\boldsymbol \mu}$ for four tomographic smoothing methods and four direct smoothing methods, and ${\boldsymbol \mu}_0$ (${\tt lake}$, ${\tt river}$ and ${\tt lake+corner}$) over 1000 runs for  combinations of $(n,N,t)$. In bold the best estimator for every setup.
  }\label{meanMSE}
\centering

\begin{tabular}{llllcccccccc}
  \hline
 &&& \multicolumn{9}{c}{Estimated MSE} \\
 \cline{5-12}
 &&&$n=$&$500$&$5000$&&$500$&$5000$&&$500$&$5000$\\
 \cline{3-12}
 N&$t$&data&estimator& \multicolumn{2}{c}{\tt lake} && \multicolumn{2}{c}{\tt river} && \multicolumn{2}{c}{\tt lake+corner} \\
 \cline{1-12}

$30$&96&tomographic&TV&{\bf 0.165}&{\bf 0.123}&&0.367&0.392&&{\bf 0.228}&{\bf 0.217}\\
     &&&splines&0.281&0.190&&0.395&{\bf 0.256}&&0.302&0.226\\
     &&&Laplacian&0.351&0.234&&0.437&0.307&&0.356&0.249\\
     &&&SVC&0.225&0.193&&{\bf 0.345}&0.309&&0.240&0.222\\
     \\&&direct&TV&0.232&0.205&&0.525&0.334&&0.331&0.229\\
     &&&GPR&0.421&0.36&&0.541&0.471&&0.439&0.385\\
     &&&Laplacian&0.549&0.411&&0.674&0.578&&0.629&0.526\\
     &&&\emph{empirical}&0.709&0.461&&0.74&0.545&&0.729&0.491\\
     \hline
     
     30&1&tomographic&TV&{\bf 0.163}&{\bf 0.120}&&0.363&0.380&&0.226&{\bf 0.208}\\
     &&&splines&0.274&0.190&&0.390&{\bf 0.255}&&0.301&0.227\\
     &&&Laplacian&0.489&0.470&&0.475&0.475&&0.477&0.470\\
     &&&SVC&0.187&0.176&&{\bf 0.295}&0.277&&{\bf 0.222}&0.212\\
     \\
     &&direct&{TV}&0.386&0.661&&0.434&0.659&&0.365&0.609\\
     &&&GPR&0.440&0.345&&0.528&0.459&&0.455&0.370\\
     &&&{Laplacian}&0.445&0.348&&0.579&0.518&&0.509&0.449\\
     &&&{empirical}&0.462&0.349&&0.545&0.463&&0.478&0.375\\
     \hline
     
     $50$&96&tomographic&TV&{\bf 0.160}&{\bf 0.130}&&0.398&{\bf 0.243}&&{\bf 0.233}&{\bf 0.159}\\
     &&&splines&0.281&0.191&&0.394&0.259&&0.301&0.231\\
     &&&Laplacian&0.412&0.327&&0.466&0.394&&0.419&0.362\\
     &&&SVC&0.23&0.197&&{\bf 0.366}&0.346&&0.249&0.232\\
     \\
     &&direct&{TV}&0.421&0.156&&0.538&0.297&&0.400&0.190\\
     &&&{GPR}&0.414&0.377&&0.533&0.482&&0.438&0.400\\
     &&&{Laplacian}&0.662&0.452&&0.751&0.598&&0.723&0.550\\
     &&&{empirical}&0.750&0.530&&0.774&0.597&&0.765&0.548\\
     \hline
     50&1&tomographic&TV&{\bf 0.148}&{\bf 0.124}&&0.393&{\bf 0.234}&&{\bf 0.225}&{\bf 0.151}\\
     &&&splines&0.277&0.191&&0.392&0.260&&0.303&0.232\\
     &&&Laplacian&0.490&0.499&&0.477&0.495&&0.472&0.503\\
     &&&SVC&0.189&0.178&&{\bf 0.318}&0.300&&0.226&0.215\\
     \\
     &&direct&{TV}&0.383&0.678&&0.420&0.677&&0.362&0.629\\
     &&&{GPR}&0.466&0.353&&0.542&0.457&&0.481&0.375\\
     &&&{Laplacian}&0.466&0.356&&0.581&0.512&&0.520&0.446\\
     &&&{empirical}&0.488&0.359&&0.561&0.462&&0.507&0.381\\
     \hline
 \end{tabular}
 \end{table}

\subsection{Coverage probability}

\begin{figure}[!t]
\centering
\includegraphics[width=5.6in]{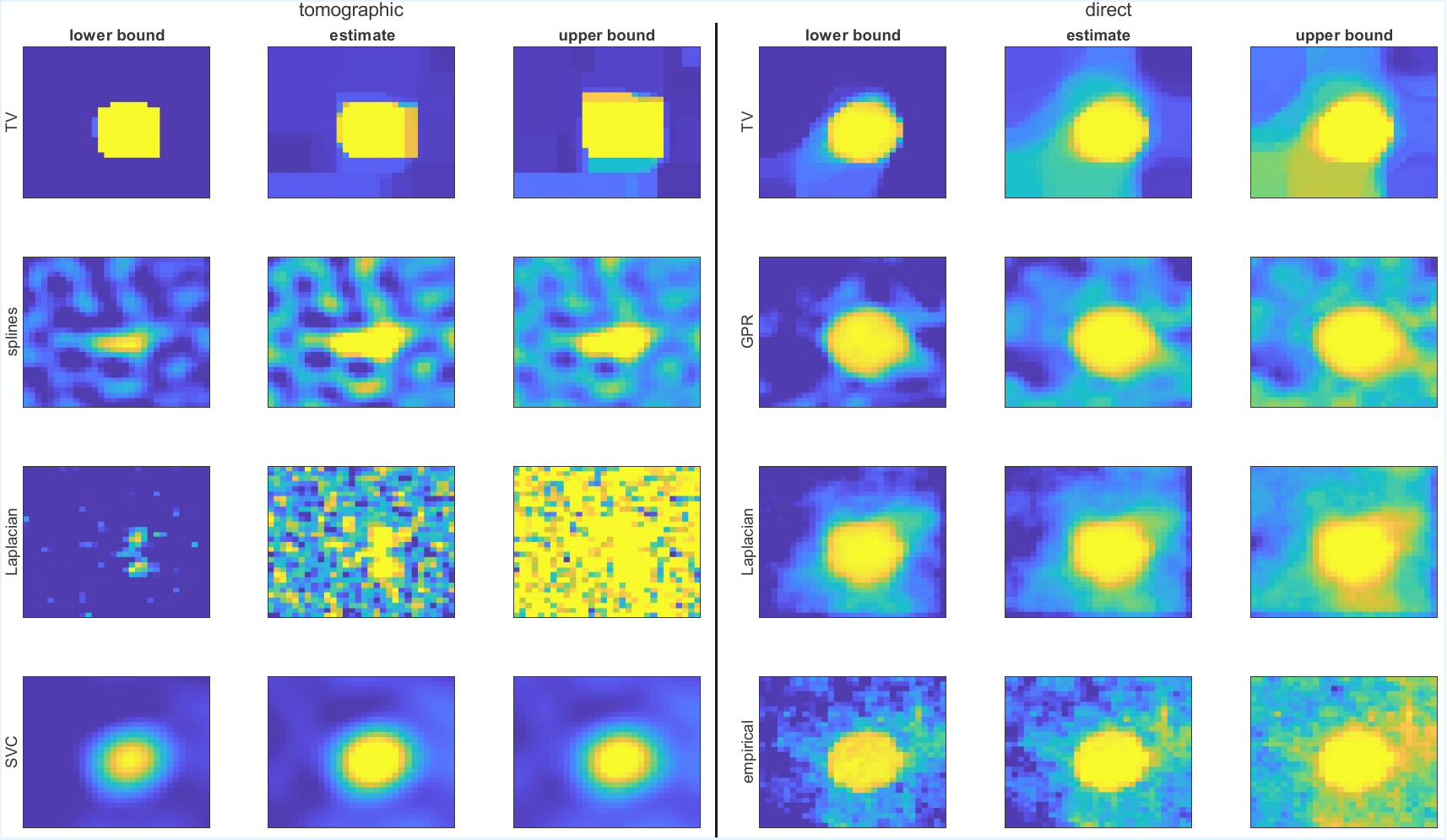}
\caption{Estimated profiles and point-wise uncertainty quantification lower and upper bounds of estimated \texttt{lake} propensity map for tomographic smoothing estimators (left) and direct smoothing estimators (right), for fixed $(n,N,t) = (500,30,1)$, $n_{\text{boot}}=5000$ and 720 sampled locations for each bootstrapped individual.}
\label{fig:coverage}
\end{figure}

We assess  coverage probability of the bootstrap-based data augmentation and uncertainty quantification method outlined in Section~\ref{subsct:uncertainty} using the parameters $(n, N, t) = (500, 30, 1)$, with $n_{\text{boot}} = 5000$ bootstrap individuals and 720 sampled locations for each bootstrap individual. A Monte Carlo simulation with 1000 iterations, summarized in Table~\ref{tab:coverage}, evaluates the average total coverage proportion of the point-wise confidence intervals. Since practitioners are mostly concerned with distinguishing high‑ and low‑risk areas, we also report the average coverage proportion separately for the low‑risk and high‑risk regions. Targeting a nominal coverage level of 0.95, the tomographic TV method exhibits close to nominal coverage for the high-risk region, while reasonable coverage for low-risk region. Other methods show notably lower coverage, with methods applied with the direct preprocessing being the worst. 
Overall, the tomographic TV method demonstrates good performance with better coverage. The diminished coverage in low‑risk regions is unsurprising, reflecting the shrinkage bias typical of penalized estimators. In contrast, the near‑nominal coverage achieved in high‑risk regions confirms that the areas of greatest concern are accurately detected. This suggests that our method provides a useful framework for practitioners to assess the significance of critical propensity hotspots. To illustrate the uncertainty measures, Figure~\ref{fig:coverage} presents typical point-wise uncertainty intervals for the eight estimators based on the \texttt{lake} simulated profile.

\begin{table}[!t]
  \caption{Results of Monte Carlo simulation reporting the averaged low-risk coverage, averaged high-risk coverage and averaged overall coverage probability (for a target value of $0.95$) of the point-wise bootstrap confidence intervals over 1000 runs for $\hat {\boldsymbol \mu}$  for four tomographic smoothing estimators and four direct smoothing estimators; for the three propensity maps ${\boldsymbol \mu}_0$ (${\tt lake}$, ${\tt river}$ and ${\tt lake+corner}$),  $(n, N, t)=(500,30,1)$, $n_{\text{boot}}=5000$, and 720 sampled locations for each bootstrap individual. In bold the closest to the nominal level.}\label{tab:coverage}
\centering
\begin{tabular}{llccccccccccc}
  \hline
 && \multicolumn{11}{c}{Average low-risk, high-risk and total coverage} \\
 \cline{3-13}
 
 \cline{3-13}
 && \multicolumn{3}{c}{\tt lake} && \multicolumn{3}{c}{\tt river} && \multicolumn{3}{c}{\tt lake+corner} \\
 \cline{3-13}
&estimator&low&high&global&&low&high&overall&&low&high&overall\\
  \hline
tomographic&TV&0.83&{\bf0.95}&0.85&&{\bf0.73}&{\bf0.99}&{\bf0.79}&&{\bf0.83}&{\bf0.93}&{\bf0.84}\\
     &splines&0.26&0.60&0.30&&0.22&0.44&0.26&&0.20&0.40&0.24\\
     &Laplacian&\bf{0.92}&0.86&{0.91}&&0.72&0.69&0.72&&0.82&0.71&0.81\\
     &SVC&0.30&0.20&0.29&&0.29&0.62&0.36&&0.38&0.21&0.35\\
     \\
     direct&{TV}&0.66&0.84&0.69&&0.55&0.97&0.63&&0.71&0.87&0.73\\
     &{GPR}&0.33&0.45&0.34&&0.29&0.45&0.32&&0.39&0.43&0.40\\
     &{Laplacian}&0.23&0.37&0.24&&0.08&0.42&0.15&&0.09&0.32&0.13\\
     &{empirical}&0.29&0.34&0.30&&0.26&0.34&0.28&&0.37&0.32&0.37\\
     \hline

 \end{tabular}
 \end{table}


\section{Application to white-tailed deer data} \label{sct:appli}

\begin{figure}[!t]
\centering
\includegraphics[width=5in]{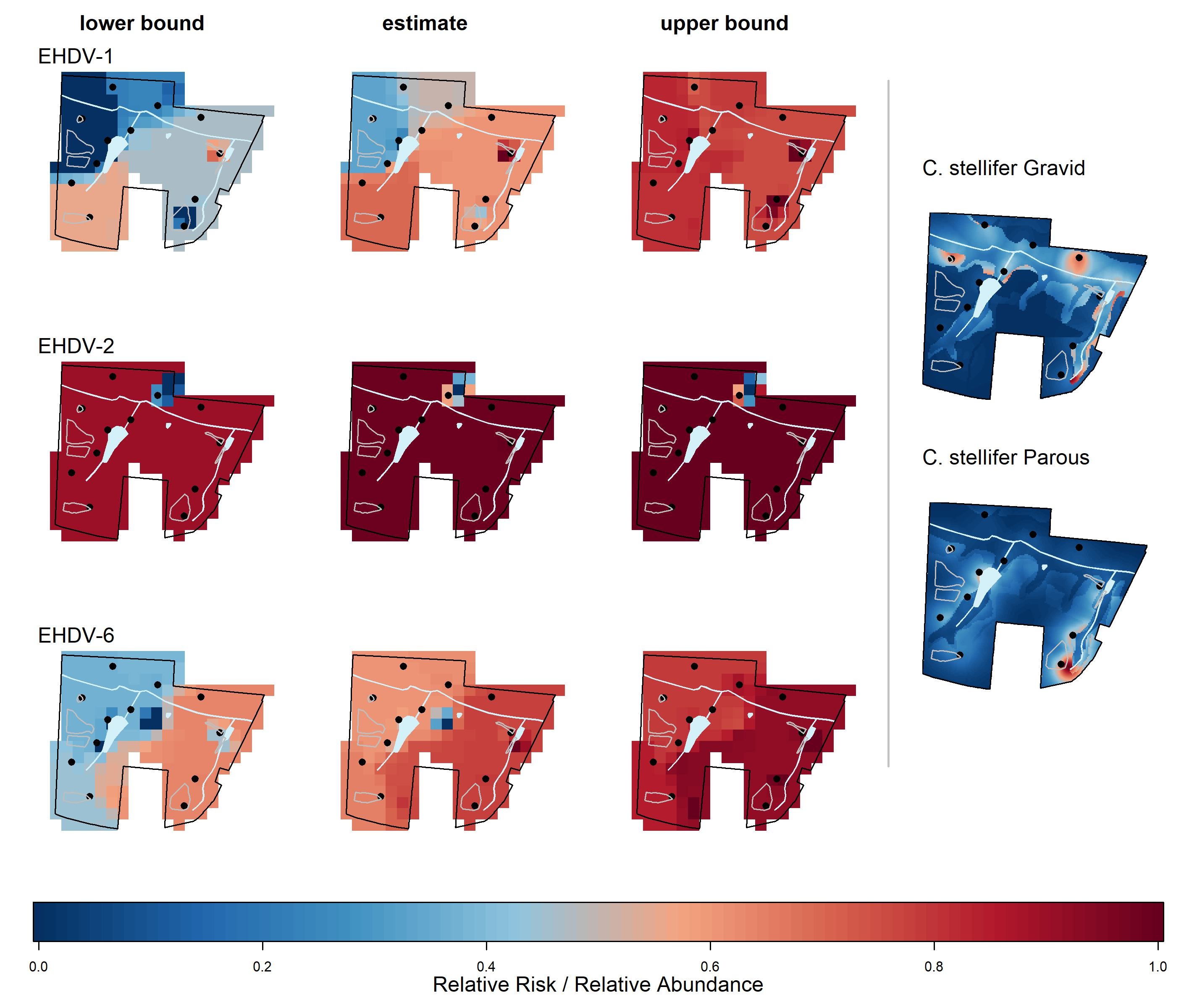}
\caption{EHDV-1, EHDV-2 and EHD-6 disease risk estimation in white tailed deer and three exotic cervid species in a Florida ranch. The left column panels show the lower bound, estimate and upper bound of the relative disease risk for each of the three serotypes evaluate based on tomographic TV estimator. Warmer colors show higher risk of getting infected. The right panels show the average relative abundance of \textit{Culicoides stellifer} in week 14 of 2015 and 2016. \textit{C.~stellifer} is a competent vector of EHDV. Each map shows feeder locations (black dots) and water sources (bright light-blue shapes and lines).}
\label{fig:deerdata}
\end{figure}

We return to the data of Section~\ref{subsct:data} and apply the proposed method, incorporating data augmentation and uncertainty quantification using $n_{boot}=600$ and $1000$ sampled locations for each bootstrap individual with 1000 runs.

\begin{figure}[!t]
\centering
\includegraphics[width=5.2in]{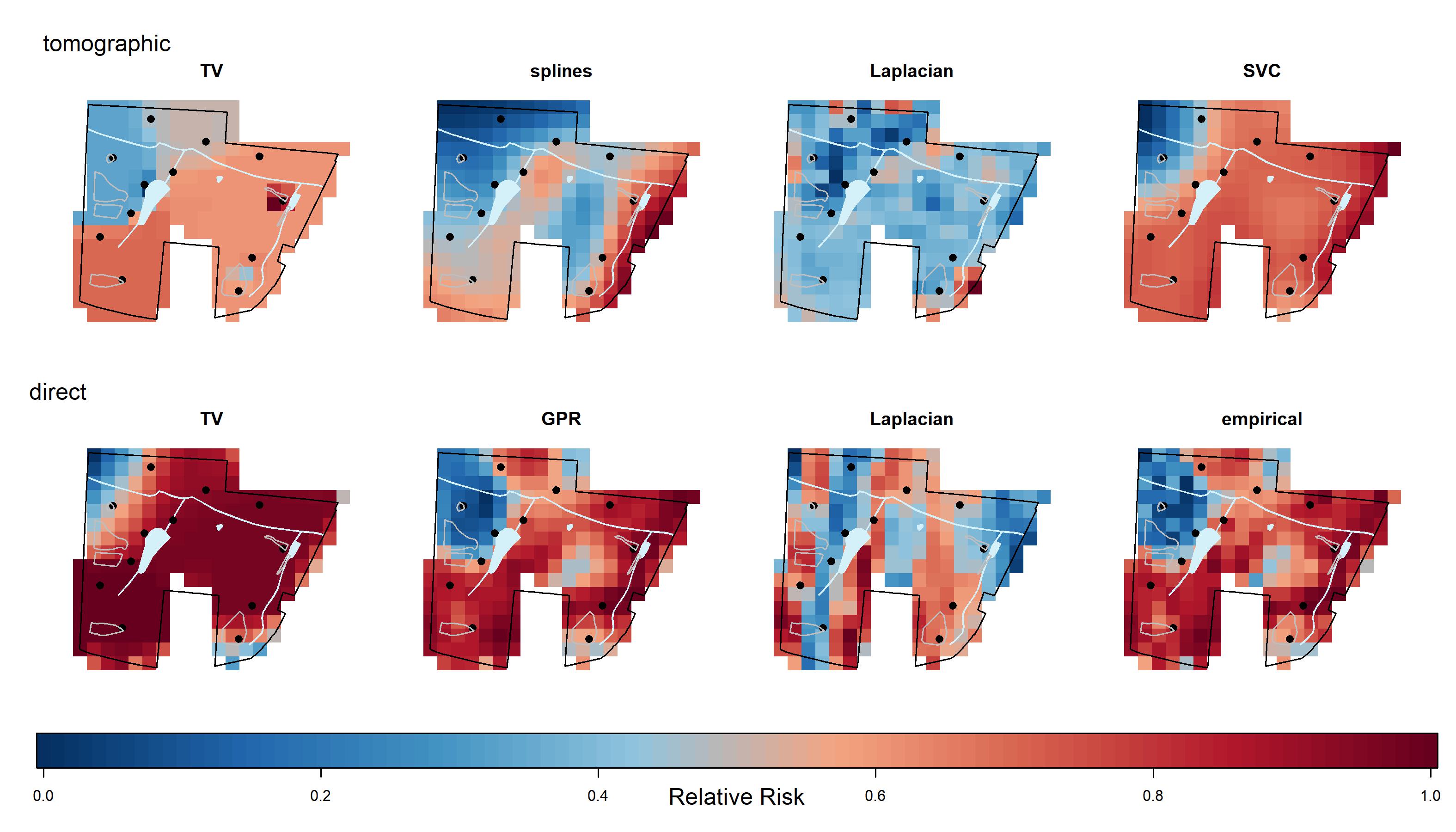}
\caption{EHDV-1 disease risk estimation in white tailed deer and three exotic cervid species in a Florida ranch, using four tomographic smoothing estimators (first row) and four direct smoothing estimators (second row). Each map shows feeder locations (black dots) and water sources (bright light-blue shapes and lines).}
\label{fig:deercompare}
\end{figure}

Figure~\ref{fig:deerdata} displays the estimated profiles obtained using our tomographic TV approach along with the associated point-wise confidence intervals (left panel). The right panel presents the observed abundance of \textit{C. stellifer}, a known vector of EHDV. Figure~\ref{fig:deercompare} compares the EHDV-1 propensities estimated with four tomographic smoothing methods and four direct smoothing methods. Figures \ref{fig:deerdata} and \ref{fig:deercompare} display feeder locations (black dots) and water sources (bright light-blue shapes and lines) for context only; these features are not included as covariates in any of the models.

Compared to the other models of disease risk estimation, tomographic TV points to smaller areas in which surveillance should be intensified for EHDV. It effectively pinpoints two specific feeders associated with EHDV-1 risk, despite having no prior knowledge of their locations. These two feeders have particularly high relative abundance of both Parous and Gravid \textit{C. stellifer} which supports the risk estimation. In comparison, most of the other models suggest that the only area where disease risk is low is the northeastern region of the ranch. Such estimates make unlikely the surveillance or control efforts since that would require controlling the abundance of \textit{C. stellifer} on  140 ha. Instead, the tomographic TV model suggests that the efforts should focus on the southeastern section and two supplementation feeders in the western portion. This reduces the initial control and surveillance area substantially. The tomographic splines and SVC models provided the most similar results to tomographic TV, albeit a larger risk area. Nonetheless, the both models correlated well with the abundance of both Gravid and Parous \textit{C. stellifer}. 

In addition to outperforming other methods, our approach offers valuable insights into the spatial distribution of disease risk across the ranch. For EHDV-1, the southwestern area was identified as having the highest disease risk (Figure~\ref{fig:deerdata}), despite the relatively low abundance of \textit{C. stellifer}. This unexpected finding suggests that midges in this region may have a higher relative prevalence of EHDV-1 compared to other areas, an insight that would not emerge from analyzing vector distribution alone.
For EHDV-2, the method indicates that nearly the entire study area is at high risk, with the exception of a small region near a feeder in the north-central part of the ranch (Figure~\ref{fig:deerdata}; \citet{benn2024culicoides}). This pattern may reflect more extensive movement of animals infected with EHDV-2 or a higher prevalence of EHDV-2 during the study period.
For EHDV-6, the eastern region of the ranch was highlighted as a high-risk area. This result aligns with the abundance of vectors in the region and underscores the importance of gravid \textit{C. stellifer} \citep{dinh2021midge}, which were more abundant than parous individuals in the northeastern part of the ranch. Interestingly, none of the models for EHDV-1 or EHDV-6 predicted high disease risk in the northeastern portion. These results can be interpreted in the context of the distribution and abundance of \textit{Culicoides} vectors across the ranch, as determined through vector collection conducted in parallel with animal tracking studies.

\section{Conclusion} \label{sct:conclu}


We introduce an innovative model that extends tomographic methodologies to estimate landscape susceptibility to various risks using GPS data, in a manner analogous to Positron Emission Tomography (PET). By adapting tomographic principles to interpret spatial data from GPS signals, our model indirectly assesses landscape-related risks, broadening tomography’s application beyond medical imaging. Our approach uniquely predicts disease transmission risk without requiring detailed knowledge of vector ecology or environmental conditions, distinguishing it from traditional models that rely on extensive host movement and vector presence data.
The sample size required and the time series length depend on the geographical extent, ecology of the disease and movement ecology of the animals.

 We validate our approach with WTD data, observing alignment with known EHDV risks and vector ecology, while providing additional insights to investigate. Comparative analyses reveal our model’s performance exceeds that of conventional spatial statistical and inverse problem-solving approaches, underscoring its robustness and adaptability. While individual recapture data is required—posing challenges for certain species—the model is applicable in contexts where movement tracking is feasible, such as avian malaria research.
 
 In addition to developing this novel approach, we propose  a selection of the total variation regularization parameter suitable for high-dimensional linear inverse problems.

\begin{acks}[Acknowledgments]
 The authors would like to thank the University of St. Gallen and the Leading House for the Latin American Region for their support through the Swiss-Latin America Seed Money Grant. First author acknowledges the support of the Natural Sciences and Engineering Research Council of Canada (NSERC); through grant DGECR-2022-04531. Data and funding for the field studies was supported by University of Florida, Institute for Food and Agricultural Sciences, Cervidae Health Research Initiative, funded through the Florida State Legislature, grant number LBR2199. The owner of the private deer ranch at the time of deer tracking enabled this study and we thank the ranch managers for their extensive logistical support. We also thank colleagues at the Florida Medical Entomology Laboratory for collecting the entomological data.
\end{acks}

\bibliographystyle{agsm}

\bibliography{article}

@string{jrssa = "Journal of the Royal Statistical Society: Series A"}

@article{threshtest2022,
    author = {Sardy, S. and Diaz-Rodriguez, J. and Giacobino, C.},
    title = {Thresholding Tests Based on Affine LASSO to Achieve Non-Asymptotic Nominal Level and High Power under Sparse and Dense Alternatives in High Dimension},
    year = {2022},
    publisher = {Elsevier Science Publishers B. V.},
    volume = {173},
    number = {C},
    journal = {Comput. Stat. Data Anal.},
    }

@article{Ahmed2015,
 author               = {Ahmed, S. and Mcinerny, G. and O'Hara, K. and Harper, R. and Salido, L. and Emmott, S. and Joppa, L.},
 journal              = {Diversity and Distributions},
 month                = {03},
 title                = {Scientists and software - surveying the species distribution modeling community},
 volume               = {21},
 year                 = {2015},
 }

@article{benn2024culicoides,
  title={Culicoides Midge Abundance across Years: Modeling Inter-Annual Variation for an Avian Feeder and a Candidate Vector of Hemorrhagic Diseases in Farmed Wildlife},
  author={Benn, Jamie S and Orange, Jeremy P and Gomez, Juan Pablo and Dinh, Emily TN and McGregor, Bethany L and Blosser, Erik M and Burkett-Cadena, Nathan D and Wisely, Samantha M and Blackburn, Jason K},
  journal={Viruses},
  volume={16},
  number={5},
  pages={766},
  year={2024},
  publisher={MDPI}
}

@book{BuhlGeer11,
 address              = {Heidelberg},
 author               = {B\"{u}hlmann, P. and van de Geer, S.},
 publisher            = {Springer},
 title                = {Statistics for {H}igh-{D}imensional {D}ata: {M}ethods, {T}heory and {A}pplications},
 year                 = {2011},
 }

@article{matheron1963principles,
  title={Principles of geostatistics},
  author={Matheron, Georges},
  journal={Economic geology},
  volume={58},
  number={8},
  pages={1246--1266},
  year={1963},
  publisher={Society of Economic Geologists}
}

@article{wood2001mgcv,
  title={mgcv: GAMs and generalized ridge regression for R},
  author={Wood, Simon N},
  journal={R news},
  volume={1},
  number={2},
  pages={20--25},
  year={2001}
}

@article{zhou2003learning,
  title={Learning with local and global consistency},
  author={Zhou, Dengyong and Bousquet, Olivier and Lal, Thomas and Weston, Jason and Sch{\"o}lkopf, Bernhard},
  journal={Advances in neural information processing systems},
  volume={16},
  year={2003}
}

@article{tikhonov1963solution,
  title={Solution of incorrectly formulated problems and the regularization method.},
  author={Tikhonov, Andrei N},
  journal={Sov Dok},
  volume={4},
  pages={1035--1038},
  year={1963}
}

@Manual{tvR,
  title        = {tvR: Total Variation Regularization},
  author       = {Minhee Lee and Taeryon Choi and Hyunwoong Ko and Youngjoo Lee},
  year         = {2022},
  note         = {R package version 1.0.3},
  url          = {https://CRAN.R-project.org/package=tvR},
}

@article{gelfand2003spatial,
  title={Spatial modeling with spatially varying coefficient processes},
  author={Gelfand, Alan E and Kim, Hyon-Jung and Sirmans, CF and Banerjee, Sudipto},
  journal={Journal of the American Statistical Association},
  volume={98},
  number={462},
  pages={387--396},
  year={2003},
  publisher={Taylor \& Francis}
}

@book{CIS-2130,
 address              = {London; New York},
 author               = {Efron, B. and Tibshirani, R.},
 publisher            = {Chapman and Hall},
 title                = {An introduction to the bootstrap},
 year                 = {1993},
 }

@article{Elith2011,
 author               = {Elith, J. and Phillips, S. and Hastie, T. and Dudik, M. and Chee, Y. and Yates, C.},
 journal              = {Diversity and Distributions},
 number               = {1},
 pages                = {43-57},
 title                = {A statistical explanation of MaxEnt for ecologists},
 volume               = {17},
 year                 = {2011},
 }

@article{Giacoetal17,
 author               = {Giacobino, C. and Sardy, S. and Diaz-Rodriguez, J. and Hengardner, N.},
 journal              = {Electronic Journal of Statistics},
 pages                = {4701--4722},
 title                = {Quantile universal threshold},
 volume               = {11},
 year                 = {2017},
 }

@article{Morris2016,
 author               = {Morris, L. and Blackburn, J.},
 journal              = {EcoHealth},
 month                = {Jun},
 number               = {2},
 pages                = {262--273},
 title                = {Predicting Disease Risk, Identifying Stakeholders, and Informing Control Strategies: A Case Study of Anthrax in Montana},
 volume               = {13},
 year                 = {2016},
 }

@article{NW72,
 author               = {Nelder, J.~A. and Wedderburn, R.~W.~M.},
 journal              = jrssa,
 number               = {3},
 pages                = {370--384},
 title                = {Generalized Linear Models},
 volume               = {135},
 year                 = {1972},
 }

@article{ROF92,
 author               = {Rudin, L.~I. and Osher, S. and Fatemi, E.},
 journal              = {Physica D},
 pages                = {259-268},
 title                = {Nonlinear total variation based noise removal algorithms},
 volume               = {60},
 year                 = {1992},
 }

@article{sardy2019,
 author               = {Sardy, S. and Monajemi, H.},
 journal              = {Journal of Computational and Graphical Statistics},
 number               = {1},
 pages                = {23-35},
 publisher            = {Taylor & Francis},
 title                = {Efficient Threshold Selection for Multivariate Total Variation Denoising},
 volume               = {28},
 year                 = {2019},
 }

@article{SAT01,
 author               = {Sardy, S. and Antoniadis, A. and Tseng, P.},
 journal              = {Journal of Computational and Graphical Statistics},
 number               = {2},
 pages                = {399--421},
 title                = {Automatic smoothing with wavelets for a wide class of distributions},
 volume               = {13},
 year                 = {2004},
 }

@article{Tibs:regr:1996,
 author               = {Tibshirani, R.},
 journal              = {Journal of the Royal Statistical Society, Series B},
 number               = {1},
 pages                = {267--288},
 title                = {Regression Shrinkage and Selection Via the Lasso},
 volume               = {58},
 year                 = {1996},
 }

@book{Wahb:spli:1990,
 address              = {Philadelphia},
 author               = {Wahba, G.},
 pages                = {169},
 publisher            = {Society for Industrial and Applied Mathematics},
 title                = {Spline Models for Observational Data},
 year                 = {1990},
 }

@article{albery2022,
  title={Fine-scale spatial patterns of wildlife disease are common and understudied},
  author={Albery, G. and Sweeny, A. and Becker, D. and Bansal, S.},
  journal={Functional Ecology},
  volume={36},
  number={1},
  pages={214--225},
  year={2022},
  publisher={Wiley Online Library}
}

@article{cauvin2020,
  title={Antibodies to epizootic hemorrhagic disease virus (EHDV) in farmed and wild Florida white-tailed deer (Odocoileus virginianus)},
  author={Cauvin, A. and Dinh, E. and Orange, J. and Shuman, R. and Blackburn, J. and Wisely, S.},
  journal={Journal of wildlife diseases},
  volume={56},
  number={1},
  pages={208--213},
  year={2020},
  publisher={Wildlife Disease Association}
}

@article{dinh2020TLoCoH,
  title={Living la Vida T-LoCoH: Site fidelity of Florida ranched and wild white-tailed deer (Odocoileus virginianus) during the epizootic hemorrhagic disease virus (EHDV) transmission period},
  author={Dinh, E. and Cauvin, A. and Orange, J. and Shuman, R. and Wisely, S. and Blackburn, J.},
  journal={Movement ecology},
  volume={8},
  pages={1--9},
  year={2020},
  publisher={Springer}
}

@article{stallknecht1996,
  title={Hemorrhagic disease in white-tailed deer in Texas: a case for enzootic stability},
  author={Stallknecht, D. and Luttrell, M. and Smith, K. and Nettles, V.},
  journal={Journal of wildlife Diseases},
  volume={32},
  number={4},
  pages={695--700},
  year={1996},
  publisher={Wildlife Disease Association}
}

@article{orange2021exotic,
  title={Evidence of epizootic hemorrhagic disease virus and bluetongue virus exposure in nonnative ruminant species in northern Florida},
  author={Orange, J. and Dinh, E. and Goodfriend, O. and Citino, S. and Wisely, S. and Blackburn, J.},
  journal={Journal of Zoo and Wildlife Medicine},
  volume={51},
  number={4},
  pages={745--751},
  year={2021},
  publisher={BioOne}
}

@article{dinh2021midge,
  title={Modeling Abundance of Culicoides stellifer, a Candidate Orbivirus Vector, Indicates Nonrandom Hemorrhagic Disease Risk for White-Tailed Deer (Odocoileus virginianus)},
  author={Dinh, E. and Gomez, J. and Orange, J. and Morris, M. and Sayler, K. and McGregor, B. and Blosser, E. and Burkett-Cadena, N. and Wisely, S. and Blackburn, J.},
  journal={Viruses},
  volume={13},
  number={7},
  pages={1328},
  year={2021},
  publisher={MDPI}
}

@article{dinh2021resource,
  title={Resource Selection by Wild and Ranched White-Tailed Deer (Odocoileus virginianus) during the Epizootic Hemorrhagic Disease Virus (EHDV) Transmission Season in Florida},
  author={Dinh, E. and Orange, J. and Peters, R. and Wisely, S. and Blackburn, J.},
  journal={Animals},
  volume={11},
  number={1},
  pages={211},
  year={2021},
  publisher={MDPI}
}

@article{orange2021homerange,
  title={Inter-annual home range fidelity of wild and ranched white-tailed deer in Florida: implications for epizootic hemorrhagic disease virus and bluetongue virus intervention},
  author={Orange, J. and Dinh, E. and Peters, R. and Wisely, S. and Blackburn, J.},
  journal={European Journal of Wildlife Research},
  volume={67},
  pages={1--8},
  year={2021},
  publisher={Springer}
}

@article{mcgregor2019host,
  title={Host use patterns of Culicoides spp. biting midges at a big game preserve in Florida, USA, and implications for the transmission of orbiviruses},
  author={McGregor, B. and Stenn, T. and Sayler, K. and Blosser, E. and Blackburn, J. and Wisely, S. and Burkett-Cadena, N.},
  journal={Medical and veterinary entomology},
  volume={33},
  number={1},
  pages={110--120},
  year={2019},
  publisher={Wiley Online Library}
}

@article{mcgregor2019field,
  title={Field data implicating Culicoides stellifer and Culicoides venustus (Diptera: Ceratopogonidae) as vectors of epizootic hemorrhagic disease virus},
  author={McGregor, B. and Sloyer, K. and Sayler, K. and Goodfriend, O. and Krauer Campos, J. and Acevedo, C. and Zhang, X. and Mathias, D. and Wisely, S. and Burkett-Cadena, N.},
  journal={Parasites \& vectors},
  volume={12},
  pages={1--13},
  year={2019},
  publisher={Springer}
}

@article{dougherty2022framework,
  title={A framework for integrating inferred movement behavior into disease risk models},
  author={Dougherty, E. and Seidel, D. and Blackburn, J. and Turner, W. and Getz, W.},
  journal={Movement ecology},
  volume={10},
  number={1},
  pages={1--15},
  year={2022},
  publisher={BioMed Central}
}

@article{tibshirani2005sparsity,
  title={Sparsity and smoothness via the fused lasso},
  author={Tibshirani, Robert and Saunders, Michael and Rosset, Saharon and Zhu, Ji and Knight, Keith},
  journal={Journal of the Royal Statistical Society: Series B (Statistical Methodology)},
  volume={67},
  number={1},
  pages={91--108},
  year={2005},
  publisher={Wiley Online Library}
}

@book{rue2005gaussian,
  title={Gaussian Markov Random Fields: Theory and Applications},
  author={Rue, H{\aa}vard and Held, Leonhard},
  year={2005},
  publisher={Chapman and Hall/CRC}
}

@article{cressie2008fixed,
  title={Fixed rank kriging for very large spatial data sets},
  author={Cressie, Noel and Johannesson, G},
  journal={Journal of the Royal Statistical Society: Series B (Statistical Methodology)},
  volume={70},
  number={1},
  pages={209--226},
  year={2008},
  publisher={Wiley Online Library}
}

@article{diaz2021nonparametric,
  title={Nonparametric Estimation of Galaxy Cluster Emissivity and Detection of Point Sources in Astrophysics With Two Lasso Penalties},
  author={Diaz-Rodriguez, Jairo and Eckert, Dominique and Monajemi, Hatef and Paltani, St{\'e}phane and Sardy, Sylvain},
  journal={Journal of the American Statistical Association},
  volume={116},
  number={535},
  pages={1088--1099},
  year={2021},
  publisher={Taylor \& Francis}
}

@article{miller2013diseases,
  title={Diseases at the livestock--wildlife interface: status, challenges, and opportunities in the United States},
  author={Miller, Ryan S and Farnsworth, Matthew L and Malmberg, Jennifer L},
  journal={Preventive veterinary medicine},
  volume={110},
  number={2},
  pages={119--132},
  year={2013},
  publisher={Elsevier}
}

@article{dougherty2018going,
  title={Going through the motions: incorporating movement analyses into disease research},
  author={Dougherty, Eric R and Seidel, Dana P and Carlson, Colin J and Spiegel, Orr and Getz, Wayne M},
  journal={Ecology letters},
  volume={21},
  number={4},
  pages={588--604},
  year={2018},
  publisher={Wiley Online Library}
}

@article{altizer2011animal,
  title={Animal migration and infectious disease risk},
  author={Altizer, Sonia and Bartel, Rebecca and Han, Barbara A},
  journal={science},
  volume={331},
  number={6015},
  pages={296--302},
  year={2011},
  publisher={American Association for the Advancement of Science}
}

@article{rayl2021elk,
  title={Elk migration influences the risk of disease spillover in the Greater Yellowstone Ecosystem},
  author={Rayl, Nathaniel D and Merkle, Jerod A and Proffitt, Kelly M and Almberg, Emily S and Jones, Jennifer D and Gude, Justin A and Cross, Paul C},
  journal={Journal of Animal Ecology},
  volume={90},
  number={5},
  pages={1264--1275},
  year={2021},
  publisher={Wiley Online Library}
}

@article{white2018disease,
  title={Disease outbreak thresholds emerge from interactions between movement behavior, landscape structure, and epidemiology},
  author={White, Lauren A and Forester, James D and Craft, Meggan E},
  journal={Proceedings of the National Academy of Sciences},
  volume={115},
  number={28},
  pages={7374--7379},
  year={2018},
  publisher={National Academy of Sciences}
}

@article{manlove2022defining,
  title={Defining an epidemiological landscape that connects movement ecology to pathogen transmission and pace-of-life},
  author={Manlove, Kezia and Wilber, Mark and White, Lauren and Bastille-Rousseau, Guillaume and Yang, Anni and Gilbertson, Marie LJ and Craft, Meggan E and Cross, Paul C and Wittemyer, George and Pepin, Kim M},
  journal={Ecology Letters},
  volume={25},
  number={8},
  pages={1760--1782},
  year={2022},
  publisher={Wiley Online Library}
}

@article{wilber2022model,
  title={A model for leveraging animal movement to understand spatio-temporal disease dynamics},
  author={Wilber, Mark Q and Yang, Anni and Boughton, Raoul and Manlove, Kezia R and Miller, Ryan S and Pepin, Kim M and Wittemyer, George},
  journal={Ecology Letters},
  volume={25},
  number={5},
  pages={1290--1304},
  year={2022},
  publisher={Wiley Online Library}
}

@article{solano2019malaria,
  title={Malaria risk assessment and mapping using satellite imagery and boosted regression trees in the Peruvian Amazon},
  author={Solano-Villarreal, Elisa and Valdivia, Walter and Pearcy, Morgan and Linard, Catherine and Pasapera-Gonzales, Jos{\'e} and Moreno-Gutierrez, Diamantina and Lejeune, Philippe and Llanos-Cuentas, Alejandro and Speybroeck, Niko and Hayette, Marie-Pierre and others},
  journal={Scientific reports},
  volume={9},
  number={1},
  pages={15173},
  year={2019},
  publisher={Nature Publishing Group UK London}
}

@article{glennon2021challenges,
  title={Challenges in modeling the emergence of novel pathogens},
  author={Glennon, Emma E and Bruijning, Marjolein and Lessler, Justin and Miller, Ian F and Rice, Benjamin L and Thompson, Robin N and Wells, Konstans and Metcalf, C Jessica E},
  journal={Epidemics},
  volume={37},
  pages={100516},
  year={2021},
  publisher={Elsevier}
}

@article{dankwa2022structural,
  title={Structural identifiability of compartmental models for infectious disease transmission is influenced by data type},
  author={Dankwa, Emmanuelle A and Brouwer, Andrew F and Donnelly, Christl A},
  journal={Epidemics},
  volume={41},
  pages={100643},
  year={2022},
  publisher={Elsevier}
}

@article{roosa2019assessing,
  title={Assessing parameter identifiability in compartmental dynamic models using a computational approach: application to infectious disease transmission models},
  author={Roosa, Kimberlyn and Chowell, Gerardo},
  journal={Theoretical Biology and Medical Modelling},
  volume={16},
  number={1},
  pages={1},
  year={2019},
  publisher={Springer}
}

@article{dekelaita2023animal,
  title={Animal movement and associated infectious disease risk in a metapopulation},
  author={Dekelaita, Daniella J and Epps, Clinton W and German, David W and Powers, Jenny G and Gonzales, Ben J and Abella-Vu, Regina K and Darby, Neal W and Hughson, Debra L and Stewart, Kelley M},
  journal={Royal Society Open Science},
  volume={10},
  number={2},
  pages={220390},
  year={2023},
  publisher={The Royal Society}
}

@article{mcduie2024mitigating,
  title={Mitigating risk: predicting h5n1 avian influenza spread with an empirical model of bird movement},
  author={McDuie, Fiona and T. Overton, Cory and A. Lorenz, Austen and L. Matchett, Elliott and L. Mott, Andrea and A. Mackell, Desmond and T. Ackerman, Joshua and De La Cruz, Susan EW and Patil, Vijay P and Prosser, Diann J and others},
  journal={Transboundary and Emerging Diseases},
  volume={2024},
  number={1},
  pages={5525298},
  year={2024},
  publisher={Wiley Online Library}
}

@article{ciss2023description,
  title={Description of the cattle and small ruminants trade network in Senegal and implication for the surveillance of animal diseases},
  author={Ciss, Mamadou and Giacomini, Alessandra and Diouf, Mame Nah{\'e} and Delabouglise, Alexis and Mesdour, Asma and Garcia Garcia, Katherin and Mu{\~n}oz, Facundo and Cardinale, Eric and Lo, Mbargou and Gaye, Adji Mar{\`e}me and others},
  journal={Transboundary and Emerging Diseases},
  volume={2023},
  number={1},
  pages={1880493},
  year={2023},
  publisher={Wiley Online Library}
}

@article{malmberg2025cross,
  title={Cross--Species Transmission at the Wildlife--Livestock Interface: A Case Study of Epidemiological Inference From Mule Deer GPS Collar Data},
  author={Malmberg, Jennifer L and Alder, Jeremy and Killion, Halcyon and Buttke, Danielle and Pepin, Kim M and Wittemyer, George},
  journal={Ecology and Evolution},
  volume={15},
  number={4},
  pages={e71182},
  year={2025},
  publisher={Wiley Online Library}
}

@article{gao2023model,
  title={Model misspecification misleads inference of the spatial dynamics of disease outbreaks},
  author={Gao, Jiansi and May, Michael R and Rannala, Bruce and Moore, Brian R},
  journal={Proceedings of the National Academy of Sciences},
  volume={120},
  number={11},
  pages={e2213913120},
  year={2023},
  publisher={National Academy of Sciences}
}

@article{akter2025conditional,
  title={Conditional logistic individual-level models of spatial infectious disease dynamics},
  author={Akter, Tahmina and Deardon, Rob},
  journal={Infectious Disease Modelling},
  volume={10},
  number={1},
  pages={268--286},
  year={2025},
  publisher={Elsevier}
}

\newpage
\begin{appendices}

\bigskip
\begin{center}
{\large\bf APPENDIX}
\end{center}

\section{Proof of Property~\ref{prop:lambda0}} 

The negative log-likelihood associated to model
$${\bf Y} \sim \operatorname{Bernoulli}(g^{-1}(X{\boldsymbol \mu}))$$ is
$$
h({\boldsymbol \mu}; {\bf y}):=-\log {\rm L} (g^{-1}(X{\boldsymbol \mu});{\bf y})=\sum_{i=1}^n y_i \log(1+\exp(-{\bf x}_i^{\rm T}{\boldsymbol \mu})+\sum_{i=1}^n (1-y_i) \log(1+\exp({\bf x}_i^{\rm T}{\boldsymbol \mu})). 
$$
Let $D$ be the $q\times p$ matrix of the $q$ finite differences such that
$\| D{\boldsymbol \mu}\|_1=\sum_{l=1}^p \sum_{l' \in \partial l} |\mu_{l}-\mu_{l'}|$. Since all differences between neighbors are penalized and there are more finite differences than elements in 
the partition of $\Omega$ (that is, $q>p$), the kernel of $D$ is spanned by the constant vector.
Moreover the cost function
\begin{equation}\label{eq:tvp}
\min_{{\boldsymbol \mu}\in{\mathbb R}^p} -\log {\rm L} (g^{-1}(X{\boldsymbol \mu});{\bf y})+\lambda \sum_{l=1}^p \sum_{l' \in \partial l} |\mu_{l}-\mu_{l'}|,
\end{equation}
 is continuous in ${\boldsymbol \mu}$ and defined on ${\mathbb R}^p$, a closed set. The function $h$ is bounded from below by zero and the penalty term tends to infinity unless ${\boldsymbol \mu}$ is in the kernel of $D$. Since all entries of X are positive (they are total times) and each row of $X$ has at least one strictly positive entry (an animal is part of the study if he spent some time in the ranch), the entries of $X{\bf 1}$ are strictly positive; consequently, it is easy to check that $\lim_{\epsilon \rightarrow \pm \infty}h(\epsilon {\bf 1}; {\bf y})=+\infty$ under the assumption the entries of ${\bf y}$ are not all ones or all zeros. 
So the cost function~\eqref{eq:tvp} is coercive. Weierstrass theorem guarantees a minimum in ${\mathbb R}^p$ for any given $\lambda>0$.

The cost function~\eqref{eq:tvp} is moreover convex,
and the Lagrange function associated to it is
$$
{\cal L}({\boldsymbol \mu}, {\boldsymbol \gamma}, {\boldsymbol \omega})=h({\boldsymbol \mu}; {\bf y})+\lambda \|{\boldsymbol \gamma} \|_1 + {\boldsymbol \omega}^{\rm T}(D{\boldsymbol \mu}-{\boldsymbol \gamma}).
$$
So the Lagrangian dual problem of~\eqref{eq:tvp} is
$$
\max_{\| {\boldsymbol \omega}\|_\infty\leq \lambda} \min_{\boldsymbol \mu}h({\boldsymbol \mu}; {\bf y})+{\boldsymbol \mu}^{\rm T}D^{\rm T}{\boldsymbol \omega} .
$$
Consequently there exists a finite $\lambda$ that leads to the constant MLE estimate, given by
$$
\min_{{\boldsymbol \omega}\in {\mathbb R}^q} \| {\boldsymbol \omega}\|_\infty \quad {\rm s.t.} \quad \nabla_{\boldsymbol \mu} h({\boldsymbol \mu}; {\bf y})+D^{\rm T}{\boldsymbol \omega} ={\bf 0},
\ {\boldsymbol \mu} =\hat \beta_0 {\bf 1},
\ \hat \beta_0=\argmin_{\beta_0} h(\beta_0 {\bf 1}; {\bf y}). 
$$

\section{Power of the TV-test} \label{subsct:test} 

\begin{figure}[!t]
\centering
\includegraphics[width=0.8\textwidth]{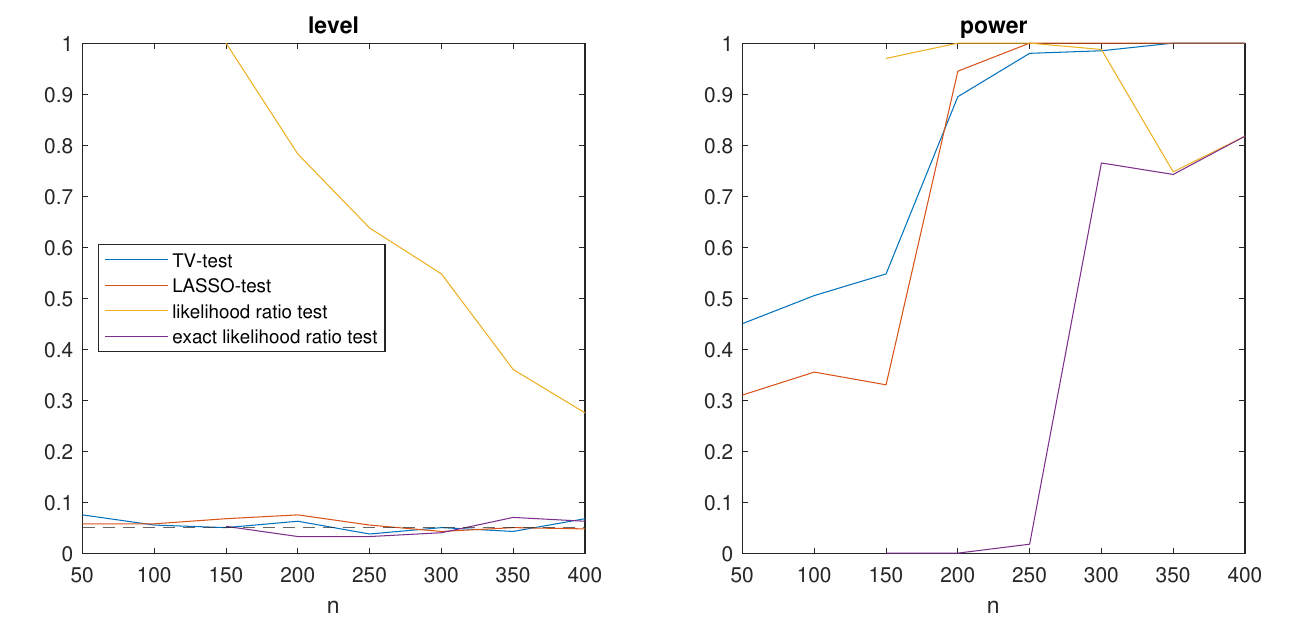}
\caption{Plots of the level  (left) and of the power (right) of four tests as a function of sample size~$n$: TV-test, LASSO-test, the likelihood ratio test based either on the asymptotic $\chi^2$ distribution or on the exact distribution empirically sampled.}
\label{fig:test}
\end{figure}

We evaluate with a Monte Carlo simulation  the power of the TV-test of Section~\ref{subsct:TVtest} as a function of the sample size $n$ under the fixed alternative hypothesis  that the propensity is the {\tt lake+corner} map.
Figure~\ref{fig:test} reports the level (left plot) and the power (right plot) of four tests: the TV-test \eqref{eq:TVtest} and the LASSO-test~\citep{threshtest2022}, the likelihood ratio test based on the asymptotic $\chi^2$ distribution, and the exact likelihood ratio test. 

As far as level is concerned (here $\alpha=0.05$), one observes that the likelihood ratio test using the asymptotic $\chi^2$ distribution is far from its nominal level when $n$ is small. Not based on asymptotic, the other three tests achieve near nominal level (thanks to a Monte Carlo simulation  to evaluate empirically the distribution of their respective test statistics under the null).

As far as power is concerned, the right plot of  Figure~\ref{fig:test} reveals that under an alternative hypothesis with spatial coherence like the  {\tt lake+corner}  map, the TV-test is more powerful than  the exact likelihood ratio test and the LASSO-test. The latter would be more powerful under a sparse alternative hypothesis since the LASSO penalty helps detect sparsity and the TV penalty helps detect spatial coherence.

\section{Data summary table} \label{app:A0}
Table~\ref{animalinfo} summarizes raw data and shows the animal tracks that were used to estimate the disease risk of three EHDV serotypes in White-tailed deer and three exotic Cervid species. We show Animal ID, sex, tracking year, start and end date of tracking period, total number of days each animal was tracked and whether the animal was seronegative (0) or seropositive (1) for each serotype at the end of the tracking period. NA values indicate animals for which serostatus was unavailable. All animals were seronegative at the beginning of the tracking period. OV indicates \textit{Odocoileus virginianus}, White-tailed deer. DD indicates \textit{Dama dama}, Fallow deer. ED indicates \textit{Elaphurus davidianus}, Pere David deer. CC indicates \textit{Cervus canadensis}, Elk deer.

\begin{table}[!ht]
\tiny
    \caption{Table summarizing raw data from White-tailed deer and three exotic Cervid species tracked to estimate the disease risk of three EHDV serotypes }\label{animalinfo}
    \centering
    \begin{tabular}{lcccccccc}
    \hline
        Accession ID & Sex & Year & Date Collared & Date End & Days & EHDV-1 & EHDV-2 & EHDV-6 \\ 
    \hline
        2015\_OV10 & M & 2015 & 5/27/15 & 11/3/15 & 160 & 1 & 1 & 0 \\ 
        2015\_OV11 & F & 2015 & 5/29/15 & 10/13/15 & 137 & 0 & NA & 0 \\  
        2015\_OV12 & M & 2015 & 5/27/15 & 9/23/15 & 119 & 1 & 1 & 0 \\  
        2015\_OV6 & M & 2015 & 5/28/15 & 10/3/15 & 128 & 1 & NA & 0 \\  
        2015\_OV7 & M & 2015 & 5/29/15 & 10/13/15 & 137 & 1 & 1 & 0 \\  
        2015\_OV8 & M & 2015 & 5/28/15 & 10/13/15 & 137 & 1 & NA & 0 \\  
        2015\_OV9 & M & 2015 & 5/27/15 & 10/3/15 & 129 & 1 & 0 & 0 \\  
        OV059 & F & 2016 & 4/13/16 & 9/21/16 & 161 & 1 & 1 & 1 \\  
        OV062 & M & 2016 & 4/13/16 & 9/21/16 & 161 & NA & 1 & 1 \\  
        OV063 & F & 2015 & 5/28/15 & 10/15/15 & 140 & 1 & NA & 0 \\  
        OV063 & F & 2016 & 4/13/16 & 10/3/16 & 173 & 1 & 1 & 1 \\  
        OV064 & F & 2016 & 4/14/16 & 9/17/16 & 156 & 1 & 1 & 1 \\  
        OV065 & M & 2015 & 5/29/15 & 10/12/15 & 136 & 1 & 1 & 0 \\  
        OV065 & M & 2016 & 4/14/16 & 9/21/16 & 160 & 1 & 1 & 1 \\  
        OV067 & M & 2016 & 4/14/16 & 9/2/16 & 141 & NA & NA & 1 \\  
        OV069 & F & 2016 & 4/14/16 & 10/4/16 & 173 & 1 & 1 & NA \\  
        OV070 & F & 2016 & 4/14/16 & 9/22/16 & 161 & 1 & 1 & 1 \\  
        OV071 & M & 2015 & 5/27/15 & 10/13/15 & 139 & 1 & NA & 1 \\  
        OV071 & M & 2016 & 4/14/16 & 9/22/16 & 160 & 1 & 1 & 1 \\  
        OV072 & M & 2016 & 4/14/16 & 9/21/16 & 160 & NA & NA & 0 \\  
        OV073 & M & 2016 & 4/15/16 & 9/23/16 & 161 & 1 & NA & 1 \\  
        OV074 & M & 2015 & 5/28/15 & 10/14/15 & 139 & 1 & 1 & 0 \\  
        OV074               & M & 2016 & 4/15/16 & 9/21/16  & 159 & 1 & 1 & 1 \\ 
        DD001               & F & 2016 & 4/15/16 & 4/3/17   & 353 & 0 & 1 & 0\\
        DD2015\_1   & M & 2015 & 5/28/15 & 10/12/15 & 137  & 1 & 1 & 0\\
        DD2015\_2   & M & 2015 & 5/29/15 & 10/14/15 & 138 & 0 & 0 & 0\\
        ED001               & F & 2016 & 4/15/16 & 9/21/16  & 159 & 1 & 1 & 0\\
        ED2015\_1   & M & 2015 & 5/29/15 &  10/12/15& 136 & 0 & 1 & 0\\
        CC2015\_1   & M & 2015 & 5/27/15 & 9/2/15   & 98 & 1 & 1 & 1\\
        CC2015\_2   & M & 2015 & 5/29/15 & 10/12/15 & 136 & 0 & 1 & 0
        \\
        \hline
    \end{tabular}
\end{table}

\end{appendices}
\end{document}